\documentclass[twocolumn,english,showpacs,amssymb,preprintnumbers,prc,floatfix]{revtex4}
\usepackage[T1]{fontenc}
\usepackage[latin1]{inputenc}
\usepackage{ae}
\usepackage{aecompl}
\usepackage{graphicx}
\usepackage{setspace}
\usepackage{rotating,epsfig,dcolumn}
\usepackage{graphicx,bm}

\usepackage{babel}
\makeatother
\begin{document}

\title{The occupancies of individual orbits and
the nuclear matrix element of the $^{76}$Ge neutrinoless $\beta\beta$ decay}

\author{J.~Men\'{e}ndez $^{*}$, A.~Poves $^{*}$, E. Caurier $^{+}$ and F. Nowacki $^{+}$}

\affiliation{({*}) Departamento de F\'{i}sica Te\'{o}rica and IFT-UAM/CSIC, Universidad Aut\'{o}noma
de Madrid, E-28049, Madrid, Spain\\
 (+) IPHC, IN2P3-CNRS/Universit\'{e} Louis Pasteur BP 28, F-67037 Strasbourg Cedex 2, France}

\date{\today{}}

\begin{abstract}
We discuss the variation of the nuclear matrix element (NME) for the neutrinoless double beta ($0\nu\beta\beta$) decay
of $^{76}$Ge when the wave functions are constrained to reproduce the experimental occupancies of the two nuclei
involved in the transition. In the Interacting Shell Model description the value of the NME is enhanced about 15\%
compared to previous calculations, whereas in the QRPA the NME's are reduced by 20\%-30\%. This diminishes 
the discrepancies between both  approaches.
In addition, we discuss effect of the short range correlations on the NME in the light of the 
recently proposed parametrizations based on a consistent renormalization of the $0\nu\beta\beta$ transition operator.

\end{abstract}


\maketitle

Very recently, there has been a strong experimental effort
to extract the occupation numbers 
of the nuclei $^{76}$Ge and $^{76}$Se \cite{Schiffer:2007no,Kay:2008po}
by accurate measurements of one nucleon transfer reactions. 
At present, both neutron and proton occupancies have been determined.
The main motivation to study these nuclei is that they are
the initial and final states of a $\beta\beta$ transition.
Therefore, we have the possibility to compare these experimental results
with the theoretical occupations and, if necessary, detect
which modifications would be required in the effective interactions
in order to obtain improved agreement with the experiment.
In principle, this would lead to more reliable results when obtaining, for instance,
the value of NME's for the $0\nu\beta\beta$ decay process.

In the case of the interacting shell model (ISM), the calculations reported
so far \cite{Caurier:2007wq,Menendez:2009dis} were performed using the gcn28.50 interaction. 
This interaction was obtained by a global fit to the region comprised by the
1$p_{3/2}$, 1$p_{1/2}$, 0$f_{5/2}$ and 0$g_{9/2}$ orbits ---$r_3g$
valence space, $r_3$ standing for the $p=3$ major oscillator shell,
with corresponding energy $E=\hbar\omega\left(p+{3/2}\right)$, except the highest $j$ orbit.
In addition, we had produced another interaction
based on gcn28.50, aimed to improve locally the  
quadrupole properties of the nuclei in the $A=76$ region.
In particular this interaction makes $^{76}$Se prolate
as suggested by the experimental data. We denote it by
rg.prolate.

 When the experimental occupation numbers were published, we decided to
compute them with the two available effective interactions, in order to
check the stability of the ISM $0\nu\beta\beta$ NME's with respect to
this property of the nuclear wave functions.

\begin{table}
\caption
{\label{cap:tab_occ76}
Proton and neutron occupation numbers of nuclei $^{76}$Ge and $^{76}$Se.
Experiment from Refs.~ \cite{Schiffer:2007no,Kay:2008po} {\it vs} theoretical
results, obtained for the gcn28.50 and rg.prolate interactions.}
\begin{center}\begin{tabular*}{\linewidth}{@{\extracolsep{\fill}}c|ccc}
\hline \hline
  & 1$p_{1/2}$+1$p_{3/2}$ & 0$f_{5/2}$ & 0$g_{9/2}$ \\
\hline
 & & Neutrons & \\
$^{76}$Ge (exp)        & 4.87$\pm$0.20 & 4.56$\pm$0.40 & 6.48$\pm$0.30 \\
$^{76}$Ge (gcn28.50)   & 5.19 & 5.02 & 5.79 \\
$^{76}$Ge (rg.prolate) & 4.83 & 4.78 & 6.39 \\
$^{76}$Se (exp)        & 4.41$\pm$0.20 & 3.83$\pm$0.40 & 5.80$\pm$0.30 \\
$^{76}$Se (gcn28.50)   & 4.86 & 4.54 & 4.60 \\
$^{76}$Se (rg.prolate) & 4.08 & 4.06 & 5.86 \\
\hline
 & & Protons & \\
$^{76}$Ge (exp)        & 1.77$\pm$0.15 & 2.04$\pm$0.25 & 0.23$\pm$0.25 \\
$^{76}$Ge (gcn28.50)   & 1.70 & 1.90 & 0.40 \\
$^{76}$Ge (rg.prolate) & 1.34 & 2.00 & 0.66 \\
$^{76}$Se (exp)        & 2.08$\pm$0.15 & 3.16$\pm$0.25 & 0.84$\pm$0.25 \\
$^{76}$Se (gcn28.50)   & 2.74 & 2.27 & 0.99 \\
$^{76}$Se (rg.prolate) & 2.12 & 2.79 & 1.08 \\
\hline
\hline
\end{tabular*}\end{center}
\end{table}

In Table~\ref{cap:tab_occ76} we compare the experimental occupancies along
with the theoretical ones obtained with both the gcn28.50 and rg.prolate interactions.
The occupancies obtained with the former are quite close to the
experimental ones, specially in the case of $^{76}$Ge.
However, for $^{76}$Se they lie somewhat further from experiment.
On the contrary, the interaction
rg.prolate produces occupancies for $^{76}$Se
which are almost perfect. The only drawback of this interaction is found on the
proton occupancies in $^{76}$Ge that slightly overfill the 0$g_{9/2}$ orbit
against the filling of the $p$ orbits.
In any case, the results obtained with both interactions compare reasonably
 well with the measured ones,
while the rg.prolate interaction can be said to fit quite successfully
the experimental numbers.

The QRPA occupancies deviate more from measurements than our ISM values.
In order to cure these discrepancies with the measured occupations,
Suhonen {\it et al.} \cite{Suhonen:2008yz} and \v{S}imkovic {\it et al.} \cite{Simkovic:2008cu}
have adjusted the parameters of their reference Woods-Saxon potential
in order to reproduce the experimental numbers.
The former do it such as to obtain agreement at BCS level while
the latter get the experimental numbers only after the QRPA correlations have been included.

\begin{figure}
\begin{center}
\includegraphics[%
  width=1.0\columnwidth,
  angle=0]{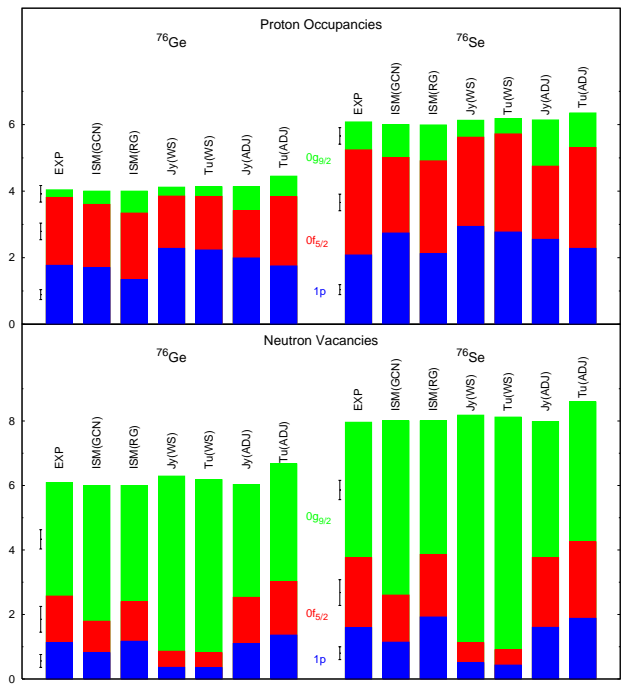}
\end{center}
\caption[Comparison between experimental and theoretical occupation numbers
for $^{76}$Ge and $^{76}$Se]
{\label{f:occ76}
Comparison between experimental and theoretical occupation numbers
for $A=76$. Experimental values from Refs.~\cite{Schiffer:2007no,Kay:2008po}.
The ISM results correspond to the gcn28.50 (GCN) and rg.prolate (RG) interactions.
The QRPA standard numbers, Tu(WS) and Jy(WS) give the occupancies at BCS level.
The QRPA occupancies with adjusted single particle energies are given at
BCS level in the case of Jy(ADJ) and at QRPA level for Tu(ADJ).
Jy and Tu results from Refs.~\cite{Suhonen:2008yz} and \cite{Simkovic:2008cu}, respectively.
The experimental error bars are also shown.}
\end{figure}

In Figure~\ref{f:occ76} we have plotted the experimental
occupancies compared to the theoretical ISM ---gcn28.50 and rg.prolate--- and QRPA values
---both original and adjusted interactions for the T\"ubingen \cite{Simkovic:2008cu}
and Jyv\"askyl\"a \cite{Suhonen:2008yz} groups.
We can observe that the amount of change in occupancies required to match
the experiment is much larger in the case of the QRPA calculations, notably for neutrons.
The effect of the new ISM interaction rg.prolate is much milder.
In the end, all final interactions are able to reproduce the experimental
occupations fairly well, with similar accuracies.

Once the interactions have been settled to give results as close as possible
to experiment, the next step is to look at the NME's.
In Table~\ref{cap:tab_nmeocc76} we have collected their values for the ISM
and QRPA with the six interactions considered, treating the SRC's by the UCOM prescription.

In the case of the Jyv\"askyl\"a group, the NME suffers a substantial
reduction of around 30\% when calculated with the adjusted interaction.
There is an effect in the same direction, whereas more moderate, present in
the T\"ubingen's results. In this case, the reduction is closer to 20\%.
The different changes are probably related to the adjustment of
experimental occupancies at BCS or QRPA level.
These modifications can be traced back to the new values of the QRPA
parameters $g_{pp}$ achieved with the modified single particle energies,
which are significantly different from those obtained with the original
single particle energies, originated from Woods-Saxon potentials.

As for the ISM, the NME obtained with the rg.prolate interaction is
enhanced with respect to the previously reported result obtained with gcn28.50.
The increase is of some 15\%. This means that our ISM is reasonably stable
when obtained with different effective interactions.
Moreover, when adjusting the interactions to agree with the measured
occupancies in $^{76}$Ge and $^{76}$Se, the difference between the ISM and QRPA
NME values diminishes. Notice however that expressing the effects in percentage may be misleading.
Indeed, in the ISM case the NME increases by 0.45 while in the two QRPA calculations the 
reductions amount to 1.25 and 0.64, respectively.

The above analysis points out the relevance of occupation numbers in order to obtain a reliable
result for the NME of the $0\nu\beta\beta$ decay. However, some caution needs to be taken regarding
this point. For instance, we have observed that, performing calculations with truncations in the
maximum seniority allowed in the wave functions ($s_m$), the occupancies obtained are essentially independent
of $s_m$, while the NME is strongly reduced when high order seniority components are allowed in the wave functions.
This can be observed in Table~\ref{cap:tab_nmeocc76.sen}.
Therefore, it is concluded that occupation numbers by themselves do not fix the NME value, even though
they are presumably necessary to get a sensible result.

\begin{table}
\caption
{\label{cap:tab_nmeocc76}
Values of the NME ($M^{0\nu\beta\beta}$) for the $^{76}$Ge  $\rightarrow$ $^{76}$Se decay
for ISM and QRPA calculations. QRPA(Jy)-WS and QRPA(Tu)-WS  are the original QRPA calculations
from Refs. \cite {Kortelainen:2007rh} ---for Jyv\"askyl\"a---
and \cite{Simkovic:2007vu} ---for T\"ubingen.
ADJ-WS are the calculations using a Woods-Saxon potential adjusted to reproduce the experimental occupancies,
obtained from Refs.~\cite{Suhonen:2008yz} (Jy) and \cite{Simkovic:2008cu} (Tu).
UCOM type SRC's are considered.
All results have $r_0=1.2$~fm and non-quenched axial coupling.
}
\begin{center}
\begin{tabular*}{\linewidth}{@{\extracolsep{\fill}}c|cccc}
\hline \hline
$M^{0\nu\beta\beta}$ & GCN & WS & RG & ADJ-WS \\
\hline
ISM & 2.81 & & 3.26 & \\
QRPA(Jy)  &  & 5.36 & & 4.11 \\
QRPA(Tu)  &  & 5.07-6.25 & & 4.59-5.44 \\
\hline
\hline
\end{tabular*}\end{center}
\end{table}

\begin{table}
\caption
{\label{cap:tab_nmeocc76.sen}
Occupancies and NME for the $^{76}$Ge  $\rightarrow$ $^{76}$Se decay
in function of the maximum seniority permitted in the wave functions, $s_m$.
UCOM type SRC's.}
\begin{center}
 \begin{tabular*}{\linewidth}{@{\extracolsep{\fill}}c|cccccccc|c}
  \hline \hline
 & \multicolumn{3}{c}{Neutrons} & &\multicolumn{3}{c}{Protons} & & NME \\
\hline
 & \multicolumn{7}{c}{$^{76}$Ge} & & \\
         & 1$p$  & 0$f_{5/2}$ & 0$g_{9/2}$  & & 1$p$ & 0$f_{5/2}$ & 0$g_{9/2}$ & &\\
\hline
$s_m=0$  & 4.8   & 5.2        & 6.1         & & 1.3  & 2.1        & 0.6        &  &\\
$s_m=4$  & 4.8   & 5.0        & 6.2         & & 1.3  & 2.0        & 0.7        &  &\\
$s_m=10$ & 4.8   & 4.8        & 6.4         & & 1.3  & 2.0        & 0.7        &  &\\
\hline
 & \multicolumn{7}{c}{$^{76}$Se} &  & \\
         & 1$p$  & 0$f_{5/2}$ & 0$g_{9/2}$ & & 1$p$ & 0$f_{5/2}$ & 0$g_{9/2}$ &  &\\
\hline
$s_m=0$  & 3.9   & 4.6        & 5.5        & & 1.8  & 3.3        & 0.9        &  &11.85 \\
$s_m=4$  & 4.3   & 4.4        & 5.3        & & 2.1  & 2.6        & 1.3        &  &7.99 \\
$s_m=14$ & 4.1   & 4.1        & 5.9        & & 2.1  & 2.8        & 1.1        &  &3.26 \\
\hline \hline
 \end{tabular*}

\end{center}
\end{table}

In the same fashion, it is interesting to look at the variation of the nuclear matrix element
of the $2\nu\beta\beta$ transition.
Since the parameter $g_{pp}$ is fixed in QRPA calculations
in order to reproduce the experimental $2\nu\beta\beta$ matrix element, in their case
no prediction is possible.
On the contrary, within the ISM we can make this comparison.
The result is that this matrix element is moderately enhanced as was the case of the $0\nu\beta\beta$
decay, changing from 0.32 MeV$^{-1}$ obtained with the gcn28.50 interaction up to 0.41 MeV$^{-1}$ when rg.prolate is employed.
They are to be compared with the experimental number $0.14\pm0.01$ MeV$^{-1}$ \cite{Barabash:2005sh}.
Beforehand, these theoretical values have to be quenched in order to take into account the valence space truncation,
which effectively quenches the Gamow-Teller strength.
This quenching factor must lie between 0.7 for $0\hbar\omega$ spaces
and 0.53 for the similar $r_4h$ valence space.
Taking 0.6 we get 0.12 and 0.15 MeV$^{-1}$, very close to the experiment.

Finally, we can study as well the NME of the same $0\nu\beta\beta$ decay in the light of
very recent treatments of short range correlations (SRC) \cite{Simkovic:2009pp,Engel:2009gb}.
These correlations haven been parametrized in the past by general prescriptions,
but now, efforts are being made in order to study them consistently, this is,
obtaining them from the regularization of the bare operator in the same way that
the bare interaction is regularized into the effective one within the nuclear medium.

This is done in Refs. \cite{Simkovic:2009pp} and \cite{Engel:2009gb}. Both obtain similar results for the
effect of short range correlations in the $0\nu\beta\beta$ process, which amount
to a  modification of less than 5\% for the former and to a reduction of about 5\% for the latter.
If we compare this quantity to the one coming from the two
standard parametrizations of SRC's for this decay, namely the
Miller-Spencer parametrization of a Jastrow type function \cite{Wu:1985xy,Miller:1975hu}
and the UCOM \cite{Kortelainen:2007rn,Roth:2005pd} approach,
it seems that the numbers obtained  with the latter method are to be considered more accurate,
being those coming from the Miller-Spencer parametrization an underestimation of the actual NME results.

\begin{table}
\caption
{\label{cap:tab_nme_src_consist}
Values of the NME for the $^{76}$Ge  $\rightarrow$ $^{76}$Se decay
for ISM interactions, using the SRC's proposed in Ref. \cite{Simkovic:2009pp}.}
 \begin{center}
  \begin{tabular*}{\linewidth}{@{\extracolsep{\fill}}c|ccc}
\hline \hline
Interaction & $M_{no\:SRC}$ & $M_{Argonne}^{0\nu\beta\beta}$ & $M_{Bonn}^{0\nu\beta\beta}$ \\
\hline
gcn28.50   &  2.89 & 2.82 & 3.00 \\
rg.prolate &  3.40 & 3.33 & 3.52 \\
\hline \hline
  \end{tabular*}
 \end{center}
\end{table}

Moreover, in Ref. \cite{Simkovic:2009pp} these effects are parametrized by
two Jastrow type functions.
Within the ISM we can take these two parametrizations and calculate the modification
that they cause on the NME's. The results are shown on Table \ref{cap:tab_nme_src_consist}.
They agree with those of Ref. \cite{Simkovic:2009pp}, showing very mild modifications of the NME's
by the SRC's, either a small increase ---in the case of the parametrization
that comes from the Bonn potential--- or decrease ---when the original potential is Argonne's.

In summary, we have calculated the orbital occupancies on nuclei $^{76}$Ge and $^{76}$Se
with our previously used interaction gcn28.50 and the new one rg.prolate, and compared
the results with the experimental numbers.
In both cases the agreement is reasonable, but the results of the rg.prolate
interaction agree extremely well with the experiment.
When we compute the NME with the new interaction, the gcn28.50 value is enhanced some 15\%,
getting closer to QRPA values.
At the same time, QRPA calculations reproducing successfully these experimental occupancies
lower their previous NME's in about 20\%, so overall, the gap between ISM and QRPA results
is reduced nearly to one half of the previous value.
This points out the importance of spectroscopic information in order to test
the validity of the nuclear matrix elements of the neutrinoless double beta decay.

This work has been supported by a grant of the Spanish Ministry of
Education and Science, FPA2007-66069, by the IN2P3-CICyT collaboration agreements,
by the Comunidad de Madrid (Spain), project HEPHACOS P-ESP-00346,
by the EU program  ILIAS N6 ENTApP WP1 and by the Spanish Consolider-Ingenio 2010 Program, CPAN (CSD2007-00042).


\end{document}